\documentstyle[epsf,12pt]{article}
\newlength{\dinwidth}
\newlength{\dinmargin}
\begin{document}
\input{psfig}
\vspace{1 cm}
\newcommand{\be}{\begin{equation}}
\newcommand{\ee}{\end{equation}}
\newcommand{\GeV}       {\mbox{${\rm GeV}$}}
\newcommand{\MeV}       {\mbox{${\rm MeV}$}}
\newcommand{\GeVsq}     {\mbox{${\rm GeV}^2$}}
\newcommand{\qsd}       {
\mbox{${Q^2}$}}
\newcommand{\x}         {\mbox{${\it x}$}}
\newcommand{\smallqsd}  {\mbox{${q^2}$}}
\newcommand{\ra}        {\mbox{$ \rightarrow $}}
\newcommand{\yjb}       {\mbox{${y_{_{JB}}}$}}
\newcommand{\gap}       {\hspace{0.5cm}}
\newcommand{\xda}       {\mbox{$x_{_{DA}}$}}
\newcommand{\qda}       {\mbox{$Q^{2}_{_{DA}}$}}
\newcommand{\gsubh}     {\mbox{$\gamma_{_{H}}$}}
\newcommand{\thee}      {\mbox{$\theta_{e}$}}
\newcommand{\mrsdm}     {\mbox{MRS~D$_-^{\prime}$\ }}
\newcommand{\mrsdz}     {\mbox{MRS~D$_0^{\prime}$\ }}
\newcommand{\sleq} {\raisebox{-.6ex}{${\textstyle\stackrel{<}{\sim}}$}}
\newcommand{\sgeq} {\raisebox{-.6ex}{${\textstyle\stackrel{>}{\sim}}$}}

\newcommand{\DS}         {\mbox{${\rm D}^\ast$}}
\newcommand{\DSp}         {\mbox{${\rm D}^{\ast +}$}}
\newcommand{\DSm}         {\mbox{${\rm D}^{\ast -}$}}
\newcommand{\DSpm}        {\mbox{${\rm D}^{\ast \pm}$}}
\newcommand{\DSt}        {\mbox{\tiny{${\rm D}^\ast$}}}
\newcommand{\DSs}        {\mbox{${\rm D}^\ast$'s}}
\newcommand{\Do}         {\mbox{${\rm D}^{0}$}}
\newcommand{\Dot}        {\mbox{\tiny{${\rm D}^{0}$}}}
\newcommand{\DECAY}      {\mbox{${\rm D}^{\ast +} \rightarrow {\rm D}^{0}
\pi^{+}
                          \rightarrow ({\rm K}^{-} \pi^{+})\pi^+$}}
\newcommand{\DECDS}       {\mbox{${\rm D}^{\ast +} \rightarrow {\rm D}^{0}
\pi^+ $ }}
\newcommand{\DECDo}       {\mbox{$ {\rm D}^{0} \rightarrow K^{-} \pi^+$}}
\newcommand{\PDS}       {\mbox{$ P_{\tiny{\DS}} $}}
\newcommand{\PDo}      {\mbox{$ P_{\tiny{\Do}} $}}

\title {
{\bf Study of \mbox{\boldmath D$^*(2010)^\pm$} Production
in \mbox{\boldmath $ep$} Collisions at HERA}
     \\
  \author{
           ZEUS Collaboration\\
                                             } }
\maketitle
\def\D*{$D^*$}
\def\g{\gamma}
\def\mev{{\ \rm MeV}}
\def\gev{{\rm\  GeV}}
\def\pT{$p_T$ }
\def\pZ{p_{\it z}}
\def\Z{{\it z}}
\def\fg#1{\noindent { figure #1}}
\def\xgm{$ x_\g^{meas}$ }
\def\xg{$ x_\g$ }
\def\delm{$\delta M$ }
\def\mkpi{$M({\rm K}^-\pi^+)$ }
\def\dkpi{${\rm D}^0 \to {\rm K}^- \pi^+$}
\def\ptds{$p_T({\rm D}^*)$}
\newcommand{\gsim}{\buildrel{>}\over{\sim}}
\vspace{5 cm}
\begin{abstract}
We report the first observation of charmed mesons with the ZEUS detector
at HERA using the decay channel ${\rm D}^{*+}\rightarrow (\Do \rightarrow
{\rm K}^-\pi^+)\pi^+$ (+ c.c.). Clear signals in the mass difference
$\Delta M$=$M$(D$^*$)--$M$(D$^0)$ as well as in the $M(K\pi)$ distribution
at the D$^0$ mass are found. The $ep$ cross section for inclusive \DSpm\
production with $Q^2<4~\GeV^2$ in the $\gamma p$ centre-of-mass energy
range $115 < W < 275$ \GeV\ has been determined to be $(32 \pm 7^{+4}_{-7}
)$ nb in the kinematic region \mbox{\{$p_T(\DS)\geq $ 1.7 \,\GeV,
$|\eta(\DS)| < 1.5 $\}}. Ex\-tra\-po\-la\-ting outside this region,
assuming a mass of the charm quark of 1.5 \GeV, we estimate the $ep$ charm
cross section to be $\sigma(e p \rightarrow c \bar{c}X ) = (0.45 \pm
0.11^{+0.37}_{-0.22}) \, \mu {\rm b} $ at \mbox{$\sqrt{s} = 296$}~\GeV\
and $\langle W \rangle = 198$ \GeV. The average $\gamma p$ charm cross
section \mbox{$\sigma(\gamma p \rightarrow c \bar{c}X )$} is found to be
\mbox{$(6.3 \pm 2.2^{+6.3}_{-3.0}) \, \mu {\rm b} $} at $\langle W \rangle
= 163$ \GeV\ and \mbox{$(16.9 \pm 5.2^{+13.9}_{-8.5}) \, \mu {\rm b} $} at
$\langle W \rangle = 243$ \GeV. The increase of the total charm
photoproduction cross section by one order of magnitude with respect to
low energy data experiments is well described by QCD NLO calculations
using singular gluon distributions in the proton.
\end{abstract}

\vspace{-21.2cm}
\begin{flushleft}
\tt DESY 95-013 \\
February 1995 \\
\end{flushleft}

\setcounter{page}{0}
\thispagestyle{empty}   
\newpage

\def\3{\ss}
\footnotesize
\renewcommand{\thepage}{\Roman{page}}
\begin{center}
\begin{large}
The ZEUS Collaboration
\end{large}
\end{center}
\noindent
M.~Derrick, D.~Krakauer, S.~Magill, D.~Mikunas, B.~Musgrave,
J.~Repond, R.~Stanek, R.L.~Talaga, H.~Zhang \\
{\it Argonne National Laboratory, Argonne, IL, USA}~$^{p}$\\[6pt]
R.~Ayad$^1$, G.~Bari, M.~Basile,
L.~Bellagamba, D.~Boscherini, A.~Bruni, G.~Bruni, P.~Bruni, G.~Cara
Romeo, G.~Castellini$^{2}$, M.~Chiarini,
L.~Cifarelli$^{3}$, F.~Cindolo, A.~Contin, M.~Corradi, I.~Gialas, \\
P.~Giusti, G.~Iacobucci, G.~Laurenti, G.~Levi, A.~Margotti,
T.~Massam, R.~Nania, C.~Nemoz, \\
F.~Palmonari, A.~Polini, G.~Sartorelli, R.~Timellini, Y.~Zamora
Garcia$^{1}$,
A.~Zichichi \\
{\it University and INFN Bologna, Bologna, Italy}~$^{f}$
% ...
\\[6pt]
\clearpage
\noindent
\hspace*{1mm}
$^{ 1}$ supported by Worldlab, Lausanne, Switzerland \\
\hspace*{1mm}
$^{ 2}$ also at IROE Florence, Italy  \\
\hspace*{1mm}
$^{ 3}$ now at Univ. of Salerno and INFN Napoli, Italy  \\
\hspace*{1mm}
% ...
$^{12}$ now with OPAL Collaboration, Faculty of Physics at Univ. of
        Freiburg \\
% ...

\begin{tabular}{lp{15cm}}
$^{a}$ &supported by the Natural Sciences and Engineering Research
         Council of Canada (NSERC) \\
$^{b}$ &supported by the FCAR of Quebec, Canada\\
$^{c}$ &supported by the German Federal Ministry for Research and
         Technology (BMFT)\\
$^{d}$ &supported by the MINERVA Gesellschaft f\"ur Forschung GmbH,
         and by the Israel Academy of Science \\
$^{e}$ &supported by the German Israeli Foundation, and
         by the Israel Academy of Science \\
$^{f}$ &supported by the Italian National Institute for Nuclear Physics
         (INFN) \\
$^{g}$ &supported by the Japanese Ministry of Education, Science and
         Culture (the Monbusho)
         and its grants for Scientific Research\\
$^{h}$ &supported by the Korean Ministry of Education and Korea Science
         and Engineering Foundation \\
$^{i}$ &supported by the Netherlands Foundation for Research on Matter
         (FOM)\\
$^{j}$ &supported by the Polish State Committee for Scientific Research
         (grant No. SPB/P3/202/93) and the Foundation for Polish-
         German Collaboration (proj. No. 506/92) \\
$^{k}$ &supported by the Polish State Committee for Scientific
         Research (grant No. PB 861/2/91 and No. 2 2372 9102,
         grant No. PB 2 2376 9102 and No. PB 2 0092 9101) \\
$^{l}$ &partially supported by the German Federal Ministry for
         Research and Technology (BMFT) \\
$^{m}$ &supported by the German Federal Ministry for Research and
         Technology (BMFT), the Volkswagen Foundation, and the Deutsche
         Forschungsgemeinschaft \\
$^{n}$ &supported by the Spanish Ministry of Education and Science
         through funds provided by CICYT \\
$^{o}$ &supported by the Particle Physics and Astronomy Research
        Council \\
$^{p}$ &supported by the US Department of Energy \\
$^{q}$ &supported by the US National Science Foundation
\end{tabular}

\newpage
\pagenumbering{arabic}
\setcounter{page}{1}
\normalsize

\section{Introduction}

In high energy $ep$ collisions at HERA the leading order QCD
contribution
to charm production is the
photon-gluon fusion (PGF) mechanism~\cite{proc1,proc2}.
In this process the photon
par\-ti\-ci\-pates as a point-like particle ({\em direct photon} process)
interacting
with a gluon from the proton and giving
a $c \bar{c}$ pair ($\gamma g \rightarrow c\bar c$).
The PGF cross section for $ep \rightarrow c\bar cX$ behaves like
\mbox{$d\sigma /dQ^2
\sim
Q^{-2}$} and is dominated by the exchange of almost real
photons ($Q^2 \approx 0$), i.e.\ by photoproduction events
where the electron is scattered by a small angle.
As a consequence, the contribution of the Deep Inelastic Scattering (DIS)
regime, $Q^{2} \gsim 4 \, \GeV^2$, where
the scattered electron is seen in the main detector,
is expected to be small compared to photoproduction.

Apart from the {\em direct} channel,
charm photoproduction at HERA can also proceed via the
{\em resolved photon} processes~\cite{owens,res},
where the photon behaves as a source
of partons which can scatter off the partons in the proton
(mainly $g g \rightarrow c\bar{c}$).
QCD-based models
predict that these types of processes
dominate over the direct processes for light quark
production~\cite{res,teojet}.
This prediction has been confirmed by measurements of the
ZEUS~\cite{zphoto,jetdir} and H1~\cite{h1} collaborations.
The predicted open charm cross section at HERA has two major uncertainties: the
mass
of the charm quark ($m_c$)~\cite{ellis,mcharm} and the structure functions of
the proton
and the photon.
Next to leading order (NLO)
corrections have been calculated and found to be
substantial~\cite{ellis,smith}.
The full NLO cross section $\sigma(ep \rightarrow c \bar{c} X)$  at HERA,
obtained using $m_c = 1.5$~\GeV\ and  the structure function
parametrisation  \mbox{MRSD$_-^\prime$}~\cite{mrsd} for the proton and
GRV HO~\cite{grv} for the photon
is predicted to be $\sim  0.9 \, \mu $b~\cite{mcharm} at $\sqrt{s} = 296$ \GeV.
A variation of $m_c$ by $\pm 0.3$ \GeV\ changes the values of
the cross sections by a factor of 2.
The predicted relative amount of the direct and resolved contributions
depends
on the photon structure function parametrisation.
For DG~\cite{dg}, GRV or
ACFGP~\cite{acf} the resolved contribution is less than 30\%, but if
the LAC1~\cite{lac} parametrisation is used the resolved contribution
can be 50\% or more and the predicted cross section
can increase by almost a factor of 2.
As a consequence, estimates of the total charm cross section
$\sigma(ep \rightarrow c \bar{c} X)$
at HERA vary between 0.3 $\mu$b and 2 $\mu$b.

We search for
open charm production at HERA with the ZEUS detector by looking for the
fragmentation products of the heavy quarks which produce
a $\DS (2010)^{\pm}$. The method relies on the tight kinematic
constraints of the decay chain\footnote{In this analysis the charge
conjugated decay
chain ${\rm D}^{*-}\rightarrow \bar{{\rm D}}^0\pi_S^{-} \rightarrow ({\rm
K}^+\pi^-)\pi_S^-$
is also included.}
\begin{equation}
{\rm D}^{*+}\rightarrow{\rm D}^0\pi_S^{+} \rightarrow ({\rm K}^-\pi^+)\pi_S^+
\label{decay}
\end{equation}
where the momentum of the pion coming from the D$^*$ (`soft pion', $\pi_S$)
is just 40 MeV in the D$^*$ rest frame. Consequently, the mass
difference \cite{PDG}
\[
   \Delta M = M ({\rm D}^0\pi_S) - M ( {\rm D}^0) = 145.42 \, \MeV
\]
can be measured much more accurately than the D$^*$ mass itself.
This low $Q$-value
of the D$^{*}\rightarrow {\rm D}^0\pi_S$ decay yields a prominent signal
in an otherwise phase space suppressed kinematic region, the threshold of
the $M({\rm K}\pi\pi_S) - M({\rm K}\pi) $ distribution.

We assume that the fraction of \DSpm\ originating from $b\bar{b}$ is
negligible~\cite{proc1}.

\section{Experimental Setup}

Data were collected during the 1993 running period,
when protons of energy $E_p=820$ \GeV\ were colliding
with electrons of $E_e=26.7$ \GeV. Collisions took place between 84 electron
and
proton bunches with typical beam currents of 10 mA. Additional unpaired bunches
of electrons and protons allowed an estimation of beam related background.

The total 1993 luminosity collected by ZEUS
was $\approx 550 $  nb$^{-1}$, of which
486 nb$^{-1}$ were used in the present work. This restricted sample contains
runs taken with stable trigger conditions and the tracking chambers operating
fully in the nominal magnetic field.

Charged particles are measured by the ZEUS inner tracking detectors,
which operate in a
magnetic field of 1.43 T provided by a thin superconducting coil.
Immediately surrounding the beampipe is the vertex detector (VXD)~\cite{vxd}
consisting of 120 radial cells, each with 12 sense wires. It
uses a slow drift velocity gas and the presently
achieved resolution is 50 $\mu$m in the central region of a cell and 150
$\mu$m near the edges.
Surrounding the VXD is
the central tracking detector (CTD) which consists of 72 cylindrical
drift chamber layers, organised into 9 `superlayers'~\cite{ctd}.
With our present calibration of the chamber,
the resolution of the CTD is around $260~\mu$m.
The resolution in transverse momentum for tracks going through all
superlayers is
$\sigma (p_T) /p_T \approx  \sqrt{ (0.005)^2\, p_{T}^2 + (0.016)^2}$ where
$p_T$ is in \GeV.
The single hit efficiency is greater than 95\%.
The efficiency for assigning hits to tracks depends on several factors:
very low
$p_T$ tracks suffer large systematic effects which reduce the
probability of hits being attached to them, and the 45 degree inclination of
the drift cells
also introduces an asymmetry between positive and
negative tracks. Nevertheless, the track reconstruction efficiency for
tracks with $p_T > 0.1$ \GeV\ is greater than 95\%.
Using the combined data from the VXD and CTD,
resolutions of $0.4$ cm in $Z$ and $0.1$ cm in radius
in the $XY$ plane\footnote{The ZEUS coordinate system
is defined as right handed with the $Z$ axis pointing in the
proton beam direction, hereafter referred to as forward, and the $X$
axis horizontal, pointing towards the centre of HERA.}
are obtained for the primary vertex reconstruction.
{}From Gaussian fits to the $Z$ vertex
distribution, the r.m.s. spread is found to be $10.5$ cm, in agreement with
the expectation from the HERA proton bunch length.

The high resolution
uranium-scintillator calorimeter (CAL)~\cite{calori} is used in the present
ana\-ly\-sis
to calculate global quantities of the events.
It covers the
polar angle range between $2.2^{\circ} < \theta <
176.5^{\circ}$, where $\theta = 0^{\circ}$ is the proton beam direction. It
consists of three parts: the rear calorimeter (RCAL), covering the backward
pseudorapidity\footnote{The pseudorapidity
$\eta$ is defined as $-\ln(\tan \frac{\theta}{2})$, where the polar
angle $\theta$ is taken with respect to the proton beam direction.
} range
($-3.4 < \eta < -0.75$);
the barrel calorimeter (BCAL) covering the central region ($-0.75 < \eta <
1.1$); and the forward calorimeter (FCAL) covering the forward region ($1.1 <
\eta < 3.8$). The calorimeter parts are subdivided into towers which
in turn are subdivided longitudinally into electromagnetic (EMC)
and hadronic (HAC) sections. The sections are subdivided into cells,
each of which is viewed by two photomultiplier tubes.
Under test beam conditions the CAL has an energy resolution, in units of
\GeV, of $\sigma_{E} = 0.35 \sqrt{E(\GeV)}$ for hadrons and
$\sigma_{E} = 0.18 \sqrt{E(\GeV)}$ for electrons.
The CAL also provides a time resolution of better than 1 ns for energy
deposits greater than 4.5~\GeV, and this timing is used for background
rejection.

We use two lead-scintillator calorimeters (LUMI)~\cite{lumi} to measure
the luminosity as well as to tag very small $Q^2$
processes. Bremsstrahlung photons emerging from the electron-proton
interaction
point (IP) at angles
below 0.5 mrad with respect to the
electron beam axis hit the photon calorimeter 107 m from the IP.
Electrons emitted from the IP at scattering angles $\pi - \theta^{\prime}_e \le
6$~mrad
and with energies
 $0.2~E_e <E_e^{\prime} < 0.9~E_e$ are deflected by beam magnets and
hit the electron calorimeter placed 35~m from the IP.

\section{Trigger Conditions}

Data are collected with a three level trigger \cite{status}. The hardwired
First
Level Trigger (FLT) is built as a
deadtime free
pipeline.
The FLT decision is based on energy deposits in the
calorimeter and luminosity detectors, on tracking information and
 on the muon chambers.
We require a logical OR of five conditions on sums
of energy in the calorimeter cells:
either the BCAL EMC energy exceeds 3.4 \GeV;
or it exceeds 2.0 \GeV,  if any track is found in the CTD;
or the RCAL EMC energy,
excluding the towers immediately adjacent to the beam pipe,
exceeds 2.0 \GeV; or it exceeds 0.5 \GeV, if any track is found in the CTD;
or the RCAL EMC energy, including the beam pipe
towers, exceeds 3.75 \GeV .

The Second Level Trigger (SLT) uses information from a subset of
detector components to separate physics events from
backgrounds. It rejects proton beam-gas events using particle
arrival times measured in the forward and rear sections of the calorimeter,
reducing the FLT trigger rate by almost an order of magnitude.

The Third Level Trigger (TLT) has available most of the event information.
It is used to apply stricter cuts on the event times and
to reject beam-halo and cosmic muons. Beam-gas interactions
are rejected by requiring:

\begin{itemize}

\item a reconstructed $Z$ vertex position within  $75$ cm of the nominal
interaction
point,

\item
$ \sum_i (E_i-p_{Zi}) \, > 3 \,  \GeV $,

\item
$ {\sum_i p_{Zi}}/{\sum_i E_i} < 0.9 $ \, ,

\end{itemize}

\noindent where the sums run over all calorimeter cells $i$ and $p_{Zi}$ is the
$Z$-component of the momentum vector assigned to each cell of energy $E_i$.

The following additional TLT requirements are made in order to
further reduce the background and the output rate:

\begin{itemize}

\item $(p_T^+)_{max} > 0.5$ \GeV\ and  $(p_T^-)_{max} > 0.5$ \GeV\
and  $(p_T^+)_{max} + (p_T^-)_{max} > 1.3$ \GeV, where
$(p_T^{\pm})_{max}$ is the track  of
positive or negative charge with the highest $p_T$,
\item transverse energy outside a cone of $10^{\circ}$
around the proton direction in excess of 5 \GeV; or ${ \sum_i (E_i-p_{Zi}) } >
15 \, \GeV$;
or ${\sum_i p_{Zi}}/{\sum_i E_i} \leq 0.8$; or an electron with energy
larger than 5~\GeV\ detected in the LUMI electron calorimeter.

\end{itemize}

\section{Kinematics}

Neutral Current  $ep$ scattering
$e~(k)~+~ p~(P) \rightarrow e~(k')~+~X$
can be described with the following kinematic variables:

\[
\sqrt{s} = \sqrt{(k~+~P)^2} \approx \sqrt{4 E_p E_e} = 296 \,\GeV \, ,
\]
the total $ep$ centre of mass energy;
\[
q^2=-Q^2=(k~-~k')^2 \, ,
\]
the four-momentum transfer squared carried by the virtual photon;
\[
y =\frac{q\cdot P}{k\cdot P} \, ,
\]
the Bjorken variable describing the energy transfer to the hadronic system; and
\[
W^2=(q+P)^2 = -Q^2+ys +m_p^2\, ,
\]
the centre of mass energy squared of the $\gamma p$ system,
where $m_p$ is the mass of the proton.

The variable $y$ can be determined to a good approximation
from the hadronic system using the Jacquet-Blondel expression~\cite{jacblo}:
\[
 y_{JB}=\frac{\sum_i (E_i - p_{Zi})}{2\cdot E_e}
\]
with the sum running over all calorimeter cells $i$ except for those associated
with the
scattered electron.

If the scattered electron is seen in the main detector (DIS events)
or in the LUMI electron calorimeter (tagged photoproduction events),
the variable $y$ can also be obtained from:
\[
y_e = 1 - \frac{E^{\prime}_e}{E_e}~\frac{1-cos \theta^{\prime}_e}{2}.
\]

\section{Monte Carlo Simulation}

 The Monte Carlo programs HERWIG\cite{herwig} and PYTHIA\cite{pythia}
are used to model the hadronic final states in $c\bar{c}$
production.
 Both generators include parton showers in the initial and final states.
Fragmentation into hadrons is simulated with the LUND string model \cite{lund}
as
implemented in JETSET\cite{jetset} in the case of PYTHIA, and
with a cluster algorithm in the case of HERWIG.
The lepton-photon vertex is modelled according
to the Weizs\"acker-Williams Approximation (WWA)~\cite{wwa}
in the case of PYTHIA,
whereas HERWIG uses exact matrix elements for the
photon-gluon fusion (PGF) component
and the  Equi\-va\-lent Photon Approximation (EPA)~\cite{epa} for resolved
processes.

For these Monte Carlo models a large number of c$ \bar{\rm c} $
events was generated,
but only those containing at least one charged D$^{*}$, decaying into
D$^0\pi_S$
with subsequent decay D$^0\rightarrow K \pi $,
were processed through the standard ZEUS detector and trigger simulation
programs and through the event reconstruction package.
With both Monte Carlo programs we have generated events with both direct and
resolved components,
setting
$m_c = 1.5 \, \GeV$. The parton densities in the proton were described by
\mbox{MRSD$_-^\prime$}
and by GRV HO in the case of the photon.
For systematic checks, we also generated events using different
parametrisations for  the proton (MRSD$_0^\prime$~\cite{mrsd},
CTEQ2M$^\prime$~\cite{cteq} and GRV HO~\cite{grvp})
and the photon (DG, ACFGP and LAC1).
To check the dependence of the results on the charm mass assumed in the Monte
Carlo simulations the whole analysis was repeated
using the default values for $m_c$
in both Monte Carlo generators (1.35 \GeV\ in PYTHIA and 1.8 \GeV\ in HERWIG).

\section{\mbox{\boldmath D$^*$} Observation}

\subsection{ Reconstruction Method }

In order to
select a kinematic
region where the efficiency of the tracking detectors is high
and systematic uncertainties are well understood, the following requirements
on the tracks are made:

\begin{itemize}

\item $p_T > 0.16 \, {\rm GeV} $;

\item $ | \eta | < 1.5 $, corresponding to $ 25^\circ \, < \, \theta \, <
155^\circ $.

\end{itemize}

Pairs of these tracks with opposite charge are combined and considered
in turn to be a kaon or a pion. The combination is accepted as a possible
\Do\ candidate if the K$\pi$ invariant mass lies in the range:
\[
1.80 \,  < M(K \pi) < 1.93 \, \GeV
\]
\noindent (the nominal value of the \Do\ mass is $1.865$ \GeV~\cite{PDG}).
To reconstruct \DS\  mesons, these \Do\ candidates are combined with an
additional
track
having opposite charge to that of the kaon. Assuming this third track to be the
soft pion,
the mass difference $\Delta M = M({\rm K} \pi \pi_S) - M({\rm K} \pi)$ is then
evaluated.

Monte Carlo studies show that after these cuts
more than $95 \% $
of the decay products of the
\Do\ satisfy $p_T > 0.5$ \GeV,
with a mean value of 1.5 \GeV. These high transverse momentum tracks have a
higher
reconstruction efficiency  and a better track  extrapolation
to the vertex.
As a consequence, the following more stringent cut is applied to them:
\[
p_T(K), p_T(\pi) > 0.5 \, {\rm GeV}.
\]

{}From Monte Carlo studies we find that the $\pi_S$
travels essentially in the same direction as the
\DS\ itself. Therefore the $|\eta| < 1.5$ cut on the single tracks limits
the $\eta ( \DS )$ range to:
\[
| \eta (\DS) | < 1.5.
\]
This cut was thus also applied.

Moreover, we have restricted our analysis to:
\[
p_T(\DS) > 1.7 \, {\rm GeV},
\]
since more than $ 95 \%$ of the
\DS 's fulfil this condition after the above cuts.

\subsection{\mbox{\boldmath $\Delta M$} and \mbox{\boldmath $M(\Do)$} Signals }

\begin{figure}
\epsfysize=16cm
\centerline{{\epsffile{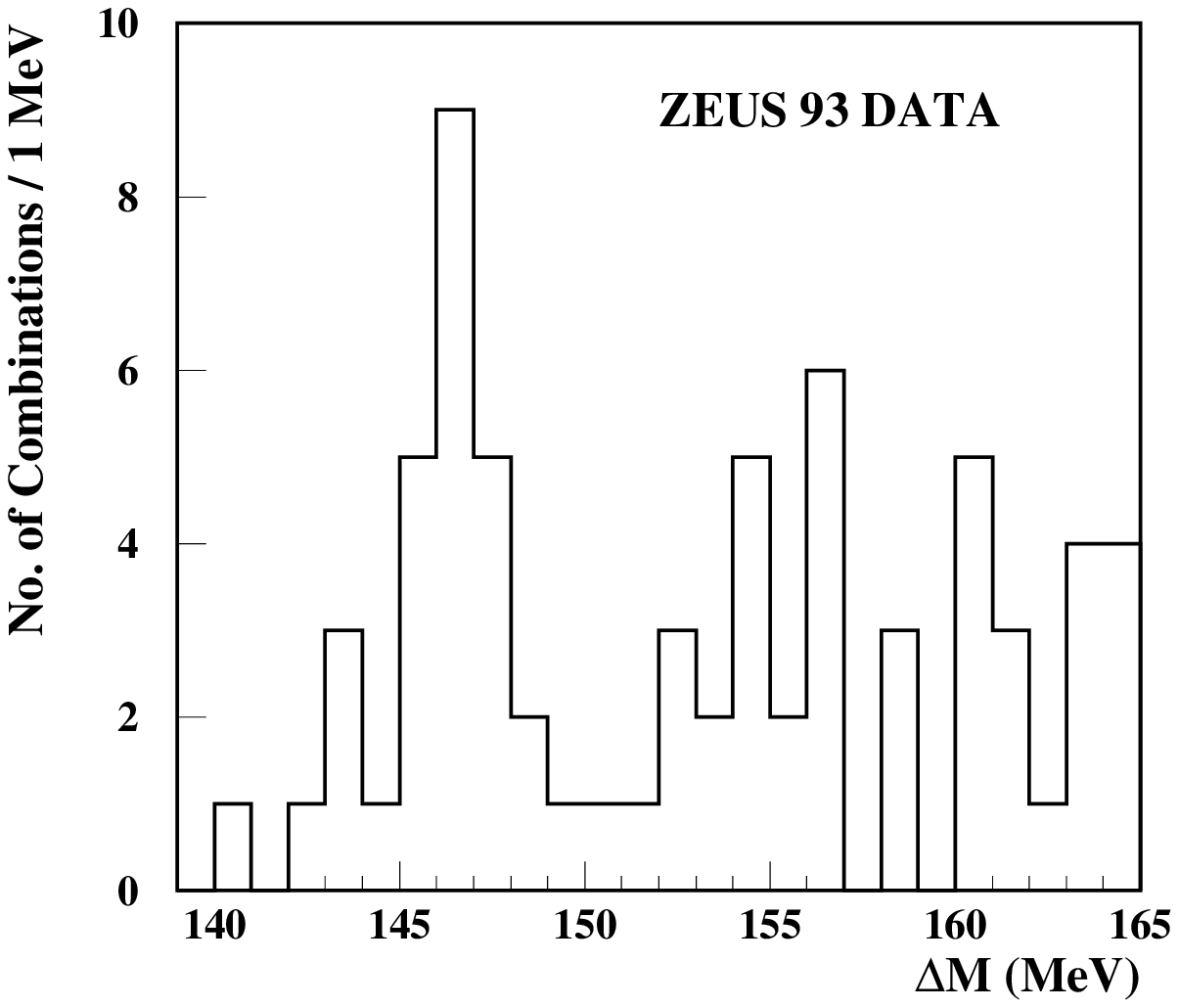}}}
\caption{\label{f:dissig} $\Delta M = M({\rm K}\pi\pi_S) - M({\rm K}\pi) $
distribution for DIS candidates : $Q^2 \ge 4 \, \GeV^2$
and $y_e<0.7$.}
\end{figure}

DIS events are defined to be those having an electron
identified in the CAL with $y_e <0.7$.
The $Q^2$ for these events is larger than 4 $\GeV^2$.
We find that 20\% of the events in which we find a \DS\ candidate fulfil this
requirement.
This relatively large fraction of DIS candidates, reproduced by the
Monte Carlo programs, is due to
the higher
trigger acceptance
for DIS events than for photoproduction events,
and to the harder $p_T(\DS)$ and $p_T(\pi_S)$ spectrum for these events.
We show the $\Delta M$ distribution for the DIS candidates in
Fig.~\ref{f:dissig}.
A clear \DSpm\ signal around the nominal value of $\Delta M$ is observed
which is evidence for \DSpm\ production at HERA in DIS with $Q^2 >  4 \,
\GeV^2$.

\begin{figure}
\epsfysize=20cm
\centerline{{\epsffile{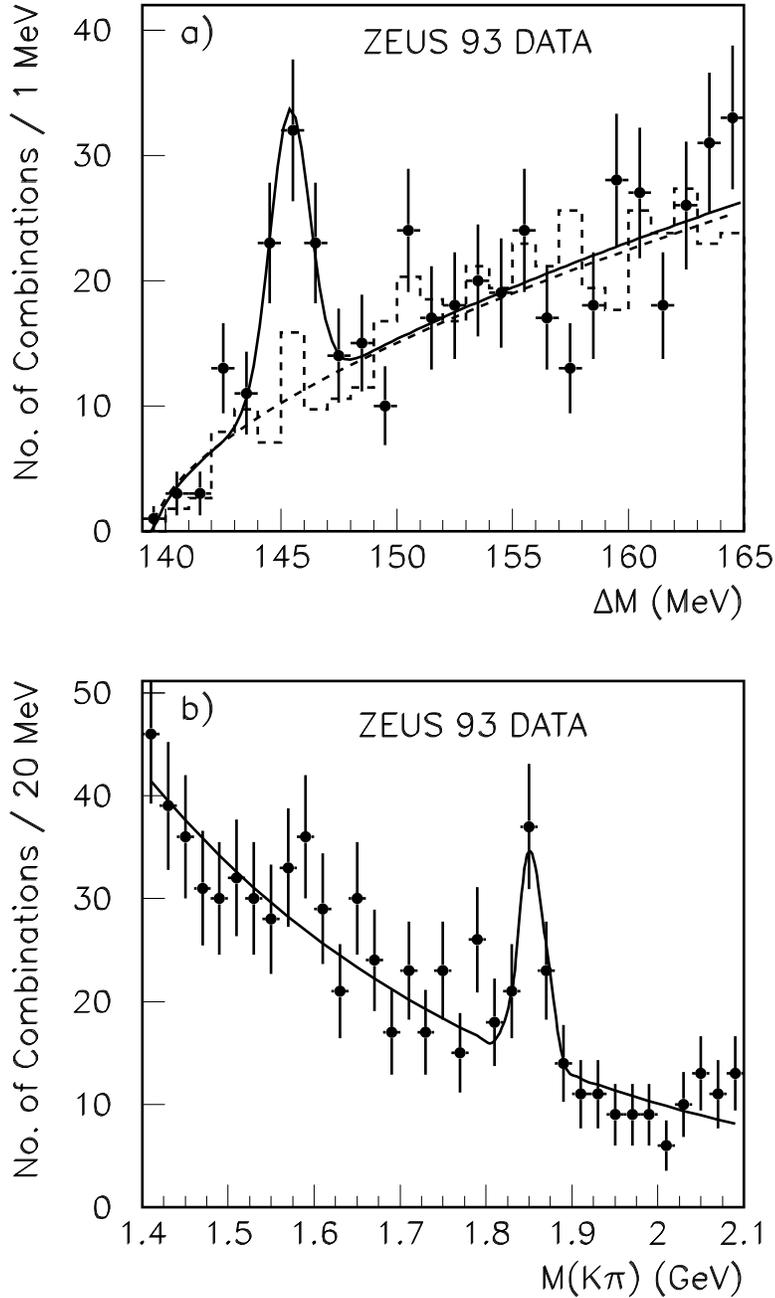}}}
\caption{\label{f:distr} a) $\Delta M$ distribution for photoproduction
events having $1.80 < M(K\pi) < 1.93$ \GeV : signal (dots)
and wrong charge combinations (dashed histogram).
The dashed line is
a fit to the wrong charge background
using the parametrisation
$A  (\Delta M - m_{\pi})^{B}$,
where $m_{\pi}$ is the  mass of the pion.
The solid line is a fit to the distribution, parametrised as a sum of
the same function for the background plus
a Gaussian for the signal.
b) $(K \pi)$ invariant mass distribution
for those candidates with $142 < \Delta M <  149$~\MeV. The fitting function is
the sum
of a Gaussian and an exponential.}
\end{figure}

The 80\% of the \DS\ candidates which are not identified as DIS events have
$Q^2 < 4 \, \GeV ^2$ and are called photoproduction events.
Of these, $27\%$ are tagged in the LUMI electron calorimeter, in agreement with
Monte Carlo
simulations and the ZEUS photoproduction measurements~\cite{zphoto}.
To reduce possible background from DIS,
where the electron is not identified,
we require \mbox{$y_{JB}<0.7$}~\cite{jetdir}.
The true $W$ and $y$ are underestimated using the Jacquet-Blondel
method. We correct the measured $W_{JB}$ with Monte Carlo
methods~\cite{incljet},
resulting in a corrected $W$ value which will be used in the following.
Comparison with the $W$ measured from events with a LUMI tag
shows that the estimated true $W$ has a systematic uncertainty of
less than $10 \%$.
The cut on $y_{JB}$ corresponds to $W < 275~\GeV$.
Furthermore we will restrict
ourselves to $W > 115$ \GeV\
where the acceptance is above 8\%.
The $\Delta M$ distribution for these photoproduction events
obtained with the set of cuts described above is shown in
Fig.~\ref{f:distr}a.
A clear peak around
$\Delta M = 145.5$  MeV is observed.

In order to check the background shape, pairs of tracks
with the same charge are selected for calculating the $K \pi$ invariant mass
(wrong charge combination method).
This distribution is fitted with the maximum-likelihood method
using the function: $A\times(\Delta M-m_{\pi})^{B}$, where $A$ and $B$ are the
free parameters of the fit.

The signal distribution is then fitted assuming this function for the
background
plus a Gaussian to parametrise the signal shape.
The corresponding fits and the normalised background are also shown in
Fig.~\ref{f:distr}a.
The background shape parameters obtained in the signal fit
agree with the values
obtained by fitting the wrong charge distribution.

We observe $ 48 \pm 11$ \DS 's above background, with
a signal to background ratio of about 1.5.
The mean value of the $\Delta M$ signal obtained from the fit is
$\Delta M \, = \, ( 145.4 \pm 0.2 ) $ \MeV,
consistent with the nominal value.
The corresponding width is
$ ( 0.9 \pm 0.2  ) $ \MeV,
in agreement with 0.7~\MeV\ obtained from Monte Carlo
simulation. The mean $W$ for these events is $\langle W \rangle = 198$ \GeV.

To check whether the \DS 's are produced according to
the decay channel (\ref{decay}), we show in Fig.~\ref{f:distr}b
the $M({\rm K} \pi)$
distribution for the events in the $\Delta M \, $ range
from  142~\MeV\ to 149~\MeV.
A clear
signal is seen around the nominal value of the \Do\ mass.
In order to fit this distribution we have used an exponential
background shape plus a Gaussian for the signal.
The number of observed \Do 's  is $43 \pm 12$, consistent with
the number of \DS 's obtained from the fit to the $\Delta M$ distribution.
We obtain a mean value of
$M(\Do) \, = \, (  1.854 \pm  0.005 ) $ \GeV,
slightly below the nominal value,
with a width of $( 18 \pm  4 )$ \MeV, consistent with the
value of 17 \MeV\ obtained from Monte Carlo simulation.

\section{Cross Sections}
\subsection{\mbox{\boldmath $ep$} Cross Section}

The cross section is obtained using the expression:
\[
\sigma  = \frac{N}{{\cal L} \times BR \times Acc} ,
\]
\noindent where $N = 48$ is the number of observed \DS 's,
${\cal L} = 486$ ${\rm nb}^{-1}$ is the integrated luminosity, and
\mbox{$BR=(2.73\pm 0.11)$\%~\cite{PDG}} is the combined branching ratio of the
decay
chain (\ref{decay}).
The acceptance $Acc$ was calculated
as the number of detected over generated D$^*\rightarrow {\rm K}\pi \pi_S$
decays  in the kinematic range chosen, using Monte Carlo methods including the
trigger simulation.
We have used the PYTHIA Monte Carlo prediction with  \mbox{MRSD$_-^\prime$/GRV
HO} structure function
parametrisations for the proton/photon to correct our data and quote
cross sections.
The overall acceptance in
the kinematic region \{$p_T(\DS)> $ 1.7 \GeV, $|\eta(\DS)| < 1.5 $\}
for the $W$ range
from 115 to 275 \GeV\ is $Acc = 11.4$\% for the above Monte Carlo program.

We describe here the checks
that are found to have a significant contribution to the systematic error.
For the acceptance, a systematic shift of $+18\%$ is estimated
using different structure function
parametrisations and  $+14\%$ using HERWIG with  \mbox{MRSD$_-^\prime$/GRV HO}
as a different Monte Carlo program.
Also, a systematic error of 8\% is found by varying
the cuts on $p_T(K,\pi)$ between 0.3~\GeV\ and
0.7 \GeV\ and on $p_T(\pi_S)$ between 0.1 \GeV\ and 0.25 \GeV.
Adding these errors in quadrature
yields a total systematic error in the acceptance
calculation of $^{+24}_{-8} \%$.
The systematic error on the number of events is 10\%, estimated
by using different background parametrisations to
fit the signals.
The systematic error of the luminosity measurement is 3.3\%.

Using the above quantities we measure an $ep$ cross section for \DSpm\
production, \\
$\sigma(ep \rightarrow \DSpm X ) \equiv \sigma(ep \rightarrow \DSp X ) +
\sigma(ep \rightarrow \DSm X )$, of:
\[
\sigma(ep \rightarrow \DSpm X )
=  32 \pm 7 (stat)^{+4}_{-7}(syst)\, \,  {\rm nb}
\]
\noindent in the kinematic region \{$p_T(\DS)> $ 1.7 \GeV,
$|\eta(\DS)|< 1.5 $\} and $115 < W < 275$ \GeV.
This cross section is valid for $Q^2 < 4 $ $\GeV^2$.
The statistical error also includes
the one due to the Monte Carlo statistics.

In order to quote a cross section for charm production we
need to correct for the fraction of events in which a charm quark pair
fragments into \DSp\ or \DSm\ as well as for the acceptance $A_{ext}$ of the
kinematic region
\mbox{\{$p_T(\DS)>1.7$ \GeV, $|\eta(\DS)| < 1.5 $\}}.
The former is $(52.0 \pm 4.2)$\%~\cite{opal}
and the latter is calculated
by using PYTHIA with \mbox{MRSD$_-^\prime$/GRV HO}
to be $A_{ext} = 13.7$\% for the region $115 < W < 275~\GeV$.
This extrapolation outside the kinematic region has a large uncertainty.
In extrapolating $p_{T}(\DS)$, the uncertainty is mainly due to the strong
dependence on the $m_{c}$ value and
for $\eta(\DS)$ it comes from the large differences between
the different structure function parametrisations in the region $|\eta(\DS)| >
1.5$.
As a consequence, the systematic error of the product $Acc \cdot A_{ext}$ is
very large.
We have fixed $m_{c}$ to 1.5 \GeV\ and quote the systematic error $\Delta(Acc
\cdot A_{ext})$
coming from
the different structure functions
and using HERWIG
(SF and MC in Table~\ref{tabla2} respectively).
Using a value of $m_c$ of 1.35 \GeV\ (1.8 \GeV) results in a shift
of $+25\% $ ($-40\% $) of the estimated cross section.

We therefore estimate the $ep$ charm production cross section
at \mbox{$\sqrt{s} = 296$}~\GeV\ for $Q^2 < 4~\GeV^2$ in the
range $115 < W <275$ \GeV\ as:
\[
\sigma(e p \rightarrow c \bar{c}X ) = 0.45 \pm 0.11^{+0.37}_{-0.22} \,\,
\mu{\rm b}.
\]

This procedure was also carried out dividing $W$ into two ranges,
$115 < W < 205$ \GeV\ and $205 < W < 275$ \GeV. The $\langle W \rangle $ for
the events
in these two ranges
were $163$ \GeV\ and $243$ \GeV\ respectively.
The error in $W$ is dominated by the systematic
uncertainty of the Jacquet-Blondel method.
The results are shown in Table~\ref{tabla2}.

\subsection{\mbox{\boldmath $\gamma p$} Cross Section }

The $\gamma p$ cross section can be obtained from the corresponding $e p$
cross section using the EPA formula:
\[
\sigma_{ep}(s) = \int_{y_{min}}^{y_{max}} dy \int_{Q^2_{min}}^{Q^2_{max}} dQ^2
\,\cdot \Phi(y,Q^2) \,
 \cdot \sigma_{\gamma^* p}(W,Q^2) ,
\]
where
\[
\Phi(y,Q^2) = \frac{\alpha}{2 \pi} \frac{1}{y Q^2}  [1+(1-y)^2- \frac{2 m_e^2
y^2}{Q^2}]
\]
is the flux of transverse photons, $Q^2_{min}=m_e^2 \frac{y^2}{1-y}$,
$Q^2_{max} =4 \,\GeV^2$ and $m_e$ is the electron mass.
Since the median \mbox{$Q^2 \approx 10^{-4}~\GeV^2$} is very small, we can
assume the photons to be
on-shell and therefore neglect the longitudinal contribution and the $Q^2$
dependence of
$\sigma_{\gamma p}$.
The $\gamma p$ cross section is then obtained as the ratio of the measured
$e p$ cross section and
the photon flux factor $\Phi$ integrated over the $Q^2$ and $y$ range
covered by the measurement.
This procedure assumes that $\sigma_{\gamma p}(W)$ is independent of $y$
in the range of the measurement. As this dependence is not known a priori,
the above calculation procedure was repeated assuming a proportional or
logarithmic
rise in $W$.
An increase of 5\% in the resulting cross section
was found at $\langle W \rangle = 163$ \GeV\
and less than 2\%
at $\langle W \rangle = 243$ \GeV.

\begin{table}[t]
\begin{center}
\begin{tabular}{||c|c|c|c|c|c|c|c|c||}\hline \hline
$\langle W \rangle$ & N & $Acc$ & $A_{ext}$ & \multicolumn{2}{|c|}{$\Delta(Acc
\cdot A_{ext})$}
                       & $\sigma (ep \rightarrow c \bar{c} X)$
                       & {\small Integrated}  & $\sigma (\gamma p \rightarrow c
\bar{c} X)$  \\
(\GeV) & & (\%) & (\%) & SF & MC & ($\mu$b)
                       & $\Phi$  & ($\mu$b)  \\
\hline \hline
$163 \pm 16$  & $21 \pm 7$  & 8.1   & 16.2 & $^{+63}_{-49} \%$ & $+54 \%$ & $
0.23 \pm 0.08^{+0.23}_{-0.11} $  & 0.0367 &  $6.3 \pm 2.2^{+6.3}_{-3.0}$  \\
\hline
$243 \pm 24$  & $28 \pm 8$  & 22.4  & 8.8  & $^{+92}_{-43} \%$ & $+30 \%$ &  $
0.21 \pm 0.06^{+0.17}_{-0.10} $  & 0.0122 & $16.9 \pm 5.2^{+13.9}_{-8.5}$  \\
\hline \hline
$198 \pm 20$  & $48\pm 11$  & 11.4  & 13.7 & $^{+76}_{-43} \%$ & $+48 \%$ &  $
0.45 \pm 0.11^{+0.37}_{-0.22} $  & 0.0488 &  $9.1 \pm 2.2^{+7.6}_{-4.4}$  \\
\hline\hline
\end{tabular}
\caption{ Acceptances and cross sections }
\label{tabla2}
\end{center}
\end{table}

\begin{figure}
\epsfysize=10cm
\centerline{{\epsffile{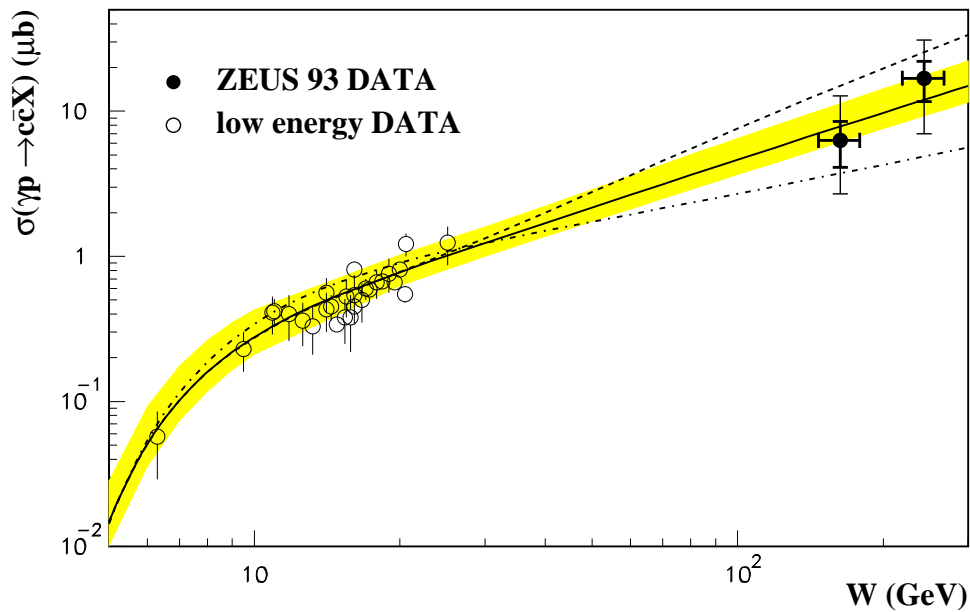}}}
\caption{ \label{f:ccplot} Total $c\bar{c}$ photoproduction cross section as a
function of $W$.
The solid dots are the ZEUS measurements and the open dots are earlier
measurements
from fixed target experiments.
The inner error bars are the statistical and the outer ones the systematic
errors.
The solid line represents the central prediction of NLO calculations with
\mbox{MRSD$_-^\prime$/GRV HO}
parametrisations of the proton/photon structure function
using $\mu_R =m_c$ (for $m_c=1.5$ \GeV) as the renormalisation scale. The
shaded band
represents the theoretical uncertainties
coming from varying this scale in the range $0.5 < \mu_R / m_c < 2$.
The dashed line represents the central prediction of
\mbox{MRSD$_-^\prime$/LAC1}
and the dash-dotted line is \mbox{MRSD$_0^\prime$/GRV HO}.
}
\end{figure}

The estimated charm photoproduction cross section
is \mbox{$(6.3 \pm 2.2^{+6.3}_{-3.0}) \, \mu {\rm b} $}
at $\langle W \rangle = 163$ \GeV\
and \mbox{$(16.9 \pm 5.2^{+13.9}_{-8.5}) \, \mu {\rm b} $}
at $\langle W \rangle = 243$ \GeV, assuming $m_c = 1.5$ \GeV\ (see
Table~\ref{tabla2}).
In Fig.~\ref{f:ccplot} we show our measurements for
the total $c\bar{c}$ photoproduction cross section as a function of $W$,
in addition to earlier measurements
in fixed target experiments~\cite{ccexp} and
NLO QCD calculations~\cite{frix2} for $m_c=1.5$ \GeV.
The solid line represents the prediction with the \mbox{MRSD$_-^\prime$/GRV HO}
structure function parametrisation for proton/photon
using $\mu_R =m_c$ as the renormalisation scale. The shaded band
represents the theoretical uncertainties
coming from varying this scale in the range $0.5 < \mu_R / m_c < 2$.
We also show the extreme predictions of \mbox{MRSD$_-^\prime$/LAC1}
(dashed line) and \mbox{MRSD$_0^\prime$/GRV HO} (dash-dotted line)
for $\mu_R =m_c$.
We note that our measurement of the
total charm photoproduction cross
section at these high $W$ values is in good
agreement with those calculated using singular gluon distributions in the
proton like
\mbox{MRSD$_-^\prime$} or GRV~\cite{grv2}.

\section{Summary}

We have observed $48$ D$^*(2010)^\pm$ mesons in the decay channel \DECAY\
(+~c.c.) in photoproduction events from $ep$ collisions at HERA.
Cross sections have been determined for these events with $Q^2 < 4 \,
\GeV^2$ and $115 < W < 275$ \GeV.
The $ep$ cross section for inclusive \DSpm\ production is found to be
$(32 \pm 7^{+4}_{-7} )$ nb in the kinematic region
\mbox{\{$p_T(\DS)> $ 1.7 \GeV,} \mbox{$|\eta(\DS)| < 1.5 $\}}.
Extrapolating outside this region and assuming a mass of the charm quark of
1.5 \GeV\
we estimate the $ep$ charm cross sections to be
$\sigma(e p \rightarrow c \bar{c}X ) = (0.45 \pm 0.11^{+0.37}_{-0.22}) \, \mu
{\rm b} $
at $\sqrt{s} = 296$~\GeV\ and $115 < W < 275$ \GeV.
The average $\gamma p$ charm cross section $\sigma(\gamma p \rightarrow c
\bar{c}X )$
is found to be
$(6.3 \pm 2.2^{+6.3}_{-3.0}) \, \mu {\rm b} $
at $\langle W \rangle = 163$ \GeV\
and $(16.9 \pm 5.2^{+13.9}_{-8.5}) \, \mu {\rm b} $
at $\langle W \rangle = 243$ \GeV.
NLO QCD calculations using
a gluon momentum density in the proton
$\sim x^{-1/2}$ at low $x$
are in good agreement with the observed increase of the
cross section by one order of magnitude when the $\gamma p$ centre-of-mass
energy
increases one order of magnitude with respect to previous low energy
experiments.

\vspace{2cm}

\noindent {\Large\bf Acknowledgements}

We thank the DESY Directorate for their strong support and
encouragement.  The remarkable achievements of the HERA machine group were
essential for the successful completion of this work,
and are gratefully appreciated.
We also gratefully acknowledge the support of the DESY computing
and network services.
We thank S.\ Frixione, M.L.\ Mangano, P.\ Nason, and G.\ Ridolfi,
for providing  us with their
results for the curves in Fig.~3.

\clearpage

\clearpage
\parskip 0mm

\clearpage

\end{document}